\documentclass{elsart}
\usepackage{natbib}
\usepackage{amsmath}
\usepackage{amssymb}
\usepackage{graphicx}

\begin{document}

\begin{frontmatter}

\title{Exponentially Localized Solutions of Mel'nikov 
Equation}

\author{C. Senthil Kumar\thanksref{label1}},
\author{ R. Radha\thanksref{label2}},
\author{M. Lakshmanan
\thanksref{label1}\corauthref{cor1}}
\ead{lakshman@cnld.bdu.ac.in}
\corauth[cor1]{Corresponding Author.  Fax:+91-431-2407093}
\address[label1]{Centre for Nonlinear Dynamics, Dept. of Physics, Bharathidasan University,
 Tiruchirapalli -620 024, India.}
\address[label2]{Dept. of Physics, Govt. College for Women, Kumbakonam - 612 001,
India.}
 
\begin{abstract}
The Mel'nikov equation is a (2+1) dimensional nonlinear evolution equation
admitting boomeron type solutions.  In this paper, after showing that it
satisfies the Painlev\'{e} property, we obtain exponentially localized dromion
type solutions from the bilinearized version which have not been reported so far.  
We also obtain more general dromion type solutions with spatially varying amplitude as well as 
induced multi-dromion solutions. 
\end{abstract}

\end{frontmatter}

\section{Introduction}
The identification of dromions which are exponentially localized solutions
in (2+1) dimensional soliton equations [1-6] has been one of the most
interesting developments in soliton theory in recent times, which has given 
a fillip to the understanding of integrable systems in (2+1) dimensions.  
Essentially, these localized solutions arise due to the presence of some
additional nonlocal terms or effective local fields associated with
``boundaries''.  Further, the advent of ``explode decay dromions'' which are 
again exponentially localized solutions with time varying amplitudes [7]
and induced dromions [6,8] using arbitrary functions of space and time
variables have set in motion the process of unearthing more and more
novel localized entities in (2+1) dimensional nonlinear systems.

\subsection{Mel'nikov Equation}
An interesting evolution equation in (2+1) dimensions which we consider here is the 
one proposed by Mel'nikov [9,10] that describes (under certain conditions) 
the interaction of two waves on the x-axis.  This equation is of the form
\begin{subequations}
\begin{eqnarray}
3u_{tt}-[u_{y}+(3u^2+u_{xx}+8\kappa|\chi|^2)_{x}]_{x} = 0, \\
i\chi_{t} = u\chi+\chi_{xx},
\end{eqnarray}
\end{subequations}

where  $u$ is the long wave amplitude (real), $\chi$ is the complex short
wave envelope, and the parameter $\kappa$ satisfies the condition $\kappa^2$ =
1. Eq. (1) may be considered either as a generalization of the K-P equation
with the addition of a complex scalar field or as a generalization of the NLS
equation with a real scalar field (after suitable interchange of coordinates $y$
and $t$).  Mel'nikov [10] has pointed out that Eq. (1)
admits boomeron type solutions, which can be realized from an asymptotic
analysis of the two soliton solution.

It is expected that the investigation of this equation may have wider
ramifications in plasma physics, nonlinear optics and hydrodynamics.   It is
this diverse presence of this equation which prompts one to make a detailed
investigation of their dynamics, particularly to identify whether localized
solutions exist in this system.  For this purpose, we first carry out a
Painlev\'e singularity structure analysis and confirm that Eq. (1) does indeed
satisfy the Painlev\'e property.  Then bilinearizing the evolution equation and
making use of certain arbitrary functions present in the solution, we obtain a
large class of exponentially localized dromion solutions. 

\section{Singularity Structure Analysis of Mel'nikov Equation}
We explore the singularity structure of Eq. (1), by rewriting
$\chi$=a and $\chi^*$=b as
\begin{subequations}
\begin{eqnarray}
3u_{tt}-u_{xy}-6u_{x}^2-6uu_{xx}-u_{xxxx}-8\kappa(a_{xx}b+2a_{x}b_{x}+ab_{xx})=0,\\
ia_{t}=ua+a_{xx},\\
-ib_{t}=ub+b_{xx}.
\end{eqnarray}
\end{subequations}
We now effect a local Laurent expansion in the neighbourhood of a
noncharacteristic singular manifold $\phi(x,y,t)$ = 0,
$\phi_{x}\neq$ 0, $\phi_{y}\neq$ 0. Assuming the leading orders of the solutions 
of Eq. (2) to have the form
\begin{equation}
u=u_{0}\phi^{\alpha},a=a_{0}\phi^{\beta},b=b_{0}\phi^{\gamma},
\end{equation}
where $u_{0}$, $a_{0}$ and $b_{0}$ are analytic functions of ($x$, $y$, $t$) and 
$\alpha$, $\beta$, $\gamma$ are integers to be determined, we now substitute 
(3) into (2) and balance the most dominant terms to get
\begin{equation}
\alpha=\beta=\gamma=-2,
\end{equation}
with the condition
\begin{equation}
\kappa a_{0}b_{0}=-9\phi_{x}^4, u_{0}=-6\phi_{x}^2.
\end{equation}
Now, considering the generalized Laurent expansion of the solutions in the 
neighbourhood of the singular manifold
\begin{subequations}
\begin{eqnarray}
u=u_{0}\phi^{\alpha}+...+u_{r}\phi^{r+\alpha}+..., \\
a=a_{0}\phi^{\beta}+...+a_{r}\phi^{r+\beta}+..., \\
b=b_{0}\phi^{\gamma}+...+b_{r}\phi^{r+\gamma}+..., 
\end{eqnarray}
\end{subequations}
the resonances (powers) at which arbitrary functions enter into (6) can be 
determined by substituting (6) into (2) and comparing the coefficients
of ($\phi^{r-6}$,$\phi^{r-4}$,$\phi^{r-4}$) to give
\begin{eqnarray}
\left(
\begin{array}{ccc}
\phi_{x}^2(r^2-5r-30)\hat{r} & 8\kappa b_{0}\hat{r} & 
8\kappa a_{0}\hat{r} \\
a_{0} & \phi_{x}^2(r^2-5r) & 0 \\
b_{0} & 0 & \phi_{x}^2(r^2-5r) 
\end{array}
\right)
\left(
\begin{array}{c}
u_{r} \\
a_{r} \\
b_{r} 
\end{array}
\right) = 0,
\end{eqnarray}
where $\hat{r}=(r-4)(r-5)$. Solving Eq. (7), one gets the resonance values as
\begin{equation}
r=-3,-1,0,4,5,5,6,8. 
\end{equation}

The resonance at $r$ = -1 naturally represents the arbitrariness of the manifold 
$\phi(x,y,t)=0$. In order to prove the existence of arbitrary functions at the other resonance
values, we now substitute the full Laurent series 
\begin{subequations}
\begin{eqnarray}
u=u_{0}\phi^{\alpha}+\sum_{r}u_{r}\phi^{r+\alpha}, \\
a=a_{0}\phi^{\beta}+\sum_{r}a_{r}\phi^{r+\beta}, \\
b=b_{0}\phi^{\gamma}+\sum_{r}b_{r}\phi^{r+\gamma} 
\end{eqnarray}
\end{subequations}
into Eq. (2).  Now collecting the coefficients of ($\phi^{-6}$,$\phi^{-4}$,$\phi^{-4}$) and 
solving them, we obtain the relations (5), implying a resonance at $r$ = 0.

Similarly collecting the following coefficients, we obtain the necessary
information about the positive resonances:

(i) coefficients of ($\phi^{-5}$,$\phi^{-3}$,$\phi^{-3}$):
\begin{eqnarray}
u_{1}=0, 
a_{1}=\frac{ia_{0}\phi_{t}}{2}-a_{0x}, 
b_{1}=\frac{-ib_{0}\phi_{t}}{2}-b_{0x}.  
\end{eqnarray}

(ii) coefficients of ($\phi^{-4}$,$\phi^{-2}$,$\phi^{-2}$): $u_{2}$, $a_{2}$,
and $b_{2}$ can be uniquely determined. 

(iii) coefficients of ($\phi^{-3}$,$\phi^{-1}$,$\phi^{-1}$): $u_{3}$, $a_{3}$, 
and $b_{3}$ can be uniquely determined. 

(iv) coefficients  of ($\phi^{-2}$,$\phi^{0}$,$\phi^{0}$): Only two equations
result for three unknowns $u_{4}$, $a_{4}$, $b_{4}$ and so one of them is
arbitrary, corresponding to a resonance at $r$ = 4.

(v) coefficients of ($\phi^{-1}$,$\phi^{1}$,$\phi^{1}$): Only $u_{5}$ is determined,
while $a_{5}$ and $b_{5}$ are arbitrary corresponding to double resonance
at $r$ = (5, 5).

(vi) coefficients of ($\phi^{0}$,$\phi^{2}$,$\phi^{2}$): Only two equations
result for three unknowns $u_{6}$, $a_{6}$, $b_{6}$ and so one of them is
arbitrary, corresponding to $r$ = 6.

(vii) coefficients of ($\phi^{1}$,$\phi^{3}$,$\phi^{3}$): $u_{7}$, $a_{7}$,
and $b_{7}$ can be determined in terms of earlier coefficients. 

(viii) coefficients  of ($\phi^{2}$,$\phi^{4}$,$\phi^{4}$): Only two equations
result for three unknowns $u_{8}$, $a_{8}$, $b_{8}$ and so one of them is
arbitrary, corresponding to $r$ = 8.

For the negative resonance $r$ = -3, following the approach of Conte, Fordy and
Pickering [11], we demand that both the solution of Mel'nikov Eq. (2) and
the solution close to it represented by a peturbation series in a small parameter
$\epsilon$ are free from movable critical manifolds.  We identify that the first 
order peturbed series does admit an
arbitrary function corresponding to a movable pole at the resonance $r$ = -3.
Consequently, for each of the eight resonances given by Eq. (8), one can
associate an arbitrary function in the solution (and close to it) without the
introduction of movable critical manifolds.  

It must be mentioned that the above system (2) admits another leading order
behaviour with $\alpha=-2$, $\beta=\gamma=-1$, $u_{0}=-2\phi_{x}^2$, $a_{0}$
and $b_{0}$ are arbitrary.  The Laurent series with the above leading order 
leads to resonances $0,0,-1,3,3,4,5,6$, corresponding to a normal branch
and the existence of sufficient number 
of arbitrary functions can be established at these resonance values.  This 
can also be verified from the corresponding bilinear form as studied by 
Grammaticos, Ramani and Hietarinta [12].   
Consequently one can be assured that the Mel'nikov equation (1) or (2) 
indeed satisfies the Painlev\'e property.

\section{Bilinearization of Mel'nikov Equation and Localized Solutions}
We next Hitora bilinearize the Mel'nikov Eq. (1) to bring out the existence
of exponentially localized solutions of Mel'nikov equation.  
Making the transformation
\begin{subequations}
\begin{eqnarray}
u  &=&  2\frac{\partial^2}{\partial x^2}\mbox{ln}F,\\
\chi  &=&  \frac{G}{F},
\end{eqnarray}
\end{subequations}
identifiable from the Painlev\'e analysis, Eq. (1) gets converted into the 
following Hirota bilinear form,
\begin{subequations}
\begin{eqnarray}
(3D_{t}^2-D_{x}D_{y}-D_{x}^4)F.F = 8\kappa|G|^2, \\
iD_{t}G.F = D_{x}^2G.F,
\end{eqnarray}
\end{subequations}
where the $D$'s are the usual bilinear operators. However, this bilinearization
have been already done  by Y. Hase etal [13] where they have given soliton 
solutions whereas we have brought out localised solutions here. For a completion, 
we proceed as follows. Introducing the series expansion, 
\begin{subequations}
\begin{eqnarray}
G &=& \epsilon g^{(1)}+\epsilon^3 g^{(3)}+..., \\
F &=& 1+\epsilon^2 f^{(2)}+\epsilon^4 f^{(4)} +... 
\end{eqnarray}
\end{subequations}
into the above bilinear form and gathering terms with various powers of the
small parameter $\epsilon$, we obtain the following set of equations,
\begin{subequations}
\begin{eqnarray}
O(\epsilon) &:& ig_{t}^{(1)} = g_{xx}^{(1)}, \\
O(\epsilon^2) &:& 3 f_{tt}^{(2)}-f_{xy}^{(2)}-f_{xxxx}^{(2)}  =  4\kappa
g^{(1)}g^{(1)*},
\end{eqnarray}
\end{subequations}
etc. Solving Eq. (14a), we can immediately write down the following solution,
\begin{equation}
g^{(1)} = \sum_{j=1}^{N}exp(\psi_{j}), \psi_{j} =
l_{j}x+m_{j}y+\omega_{j}t+\psi_{j}^{(0)}, i\omega_{j}=l_{j}^2,
\end{equation}
where the spectral parameters $l_{j}, m_{j}, \omega_{j}$ and $\psi_{j}^{(0)}$ are 
all complex. Confining to $N$ = 1 in Eq. (15) and substituting (15) into 
(14b), we obtain
\begin{equation}
f^{(2)} = \mbox{exp}(\psi_{1}+\psi_{1}^*+2A),\; \mbox{exp}(2A) = {\kappa\over
{(3\omega_{1R}^2-l_{1R}m_{1R}-4l_{1R}^4)}}.
\end{equation}
Here $l_{1}=l_{1R}+il_{1I}$, $m_{1}=m_{1R}+im_{1I}$ and $\omega_{1}=
\omega_{1R}+i\omega_{1I}$ and also $i\omega_{1}=l_{1}^2$.
Choosing $g^{(2j+1)}=0, f^{(2j)}=0$, for
$j$ $>$ 1, in Eq. (13) and  using Eqs. (15) and (16) alongwith 
the transformation (11), the physical field $u$ and the potential $\chi$ 
can be easily seen to be driven by the envelope soliton and pulse soliton 
respectively as
\begin{subequations}
\begin{eqnarray}
\chi &=& \left(\frac{\sqrt{l_{1R}}(12l_{1R}l_{1I}^2-m_{1R}-4l_{1R}^3)^
{\frac{1}{2}}}{2\sqrt\kappa}\right)
\mbox{sech}(\psi_{1R}+A)e^{i\psi_{1I}}, \\
u &=& 2l_{1R}^2\mbox{sech}^2(\psi_{1R}+A). 
\end{eqnarray}
\end{subequations}
where $\psi_{1R}=l_{1R}x+m_{1R}y+2l_{1R}l_{1I}t+\psi_{1R}^{(0)}$ and 
$\psi_{1I}=l_{1I}x+m_{1I}y-(l_{1R}^2-l_{1I}^2)t+\psi_{1I}^{(0)}$.  One can proceed further in 
the standard way to obtain higher order soliton solutions also.

\subsection{Dromions}
Looking at the above solutions (17), we realize the fact that as the parameter
$l_{1R}$ $\rightarrow$ 0, both $u$ and $\chi$ vanish. But, when $m_{1R}$ 
$\rightarrow$ $12l_{1R}l_{1I}^2-4l_{1R}^3$, the potential $\chi$ vanishes, 
whereas the physical field $u$ survives and is driven by a ghost soliton 
of the form,
\begin{equation}
u=2l_{1R}^2\mbox{sech}^2(\psi_{1R}+c), 
\end{equation}
where $c$ is a new constant. This predicts the existence of
exponentially localized solution for the complex field variable $\chi$
in Eq. (1).

\subsubsection{(1,1) Dromion}
To generate a (1,1) dromion, we now make the ansatz
\begin{equation}
F = 1+e^{\tilde{\psi}_{1}+\tilde{\psi}_{1}^*}+e^{\tilde{\psi}_{2}+\tilde{\psi}_{2}^*}
+K e^{\tilde{\psi}_{1}+\tilde{\psi}_{1}^*+\tilde{\psi}_{2}+\tilde{\psi}_{2}^*}, \\
\end{equation}
where
\begin{subequations}
\begin{eqnarray}
\tilde{\psi}_{1} &=& px+\omega t, \\
\tilde{\psi}_{2} &=& qy. 
\end{eqnarray}
\end{subequations}
Here $K$ is a real constant and $p, \omega$ and $q$ are complex constants.
Substituting (19) in (12a), we obtain
\begin{subequations}
\begin{eqnarray}
G = \rho e^{\tilde{\psi}_{1}+\tilde{\psi}_{2}}, \\
4\kappa |\rho|^2  =  (p+p^*)(q+q^*)(1-K) 
\end{eqnarray}
\end{subequations}
for the parametric choice $3\omega_{R}^2=4p_{R}^4$. Substituting (19) and (21)
in (12b), we find that $\omega=-ip^2$, where $p_{R}=\pm\sqrt{3}p_{I}$, 
$p_{R}$ and $p_{I}$ are the real and imaginary parts, respectively of $p$.
Hence, the exponentially localized solution with one bound state for the 
potential field $\chi$ for the above choice of $\omega$ and $p$ takes the form
\begin{equation}
\chi = {\rho e^{\tilde{\psi}_{1}+\tilde{\psi}_{2}}\over{1+e^{\tilde{\psi}_{1}
+{\tilde{\psi}_{1}}^*}+e^{\tilde{\psi}_{2}+{\tilde{\psi}_{2}^*}}
+K e^{\tilde{\psi}_{1}+{\tilde{\psi}_{1}}^*+\tilde{\psi}_{2}+{\tilde{\psi}_{2}}^*}}},\\
\end{equation}
while the scalar field has the form 
\begin{equation}
u = {4p_{R}^2(1+e^{\tilde{\psi}_{2}+{\tilde{\psi}_{2}}^*})
(e^{\tilde{\psi}_{1}+{\tilde{\psi}_{1}}^*}+Ke^{\tilde{\psi}_{1}+{\psi_{1}}^*
+\tilde{\psi}_{2}
+{\tilde{\psi}_{2}}^*})\over{(1+e^{\tilde{\psi}_{1}+{\tilde{\psi}_{1}}^*}
+e^{\tilde{\psi}_{2}+{\tilde{\psi}_{2}}^*}+Ke^{\tilde{\psi}_{1}+{\tilde{\psi}_{1}}^*+\tilde{\psi}_{2}
+{\tilde{\psi}_{2}}^*})^2}}, 
\end{equation}
which is always bounded, but localized everywhere except in the neighbourhood 
of the line $\tilde{\psi}_{1}+\tilde{\psi}_{1}^*=0$ in the $(x,y)$ plane. A snapshot of 
\begin{figure}[!ht]
\begin{center}
\includegraphics[width=0.75\linewidth]{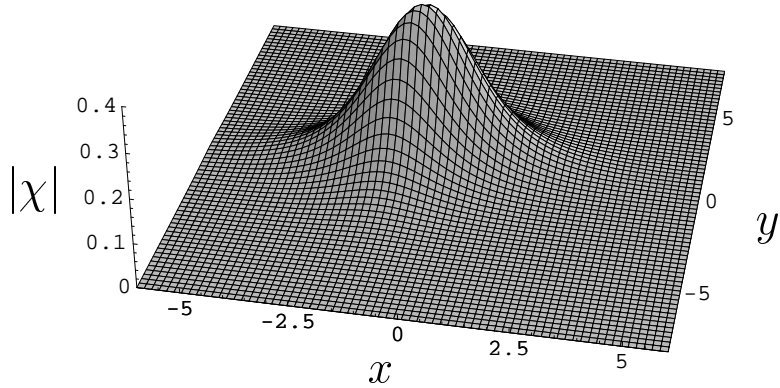}
\end{center}
\caption{Snapshot of the (1,1) dromion solution of the Mel'nikov equation 
(see Eq.~(22))}
\end{figure}
the (1,1) dromion solution  for the magnitude of the potential field $\chi$
is shown in Fig. 1.  One can proceed to find multi-dromion solutions also,
generalizing the above (1,1) dromions.

\subsubsection{Dromions with Spatially Varying Amplitude}
It can be seen from Eq. (14a) that the differential equation for $g$ involves only
the variables $t$ and $x$.  Hence, an arbitrary function of $y$ can also enter 
into its solution so that the most general form of it can be given as  
\begin{equation}
g^{(1)} = \sum_{j=1}^{N}exp(\tilde{\psi}_{j}), \tilde{\psi}_{j} =
l_{j}x+f_{j}(y)+w_{j}t+\psi_{j}^{(0)}, iw_{j}=l_{j}^2,
\end{equation}
where $f_{j}(y)$s are arbitrary functions of $y$.
This fact can be harnessed in a suitable way to construct
a more general class of localized solutions.  Following the above procedure
to derive the (1,1) dromion solution (22), one can easily 
obtain the generalized dromion solution involving arbitrary function of 
$y$ in the same form as Eq. (22) except that $\tilde{\psi}_{2}$ is now given by 
\begin{equation}
\tilde{\psi}_{2} = f(y),
\end{equation}
instead of (20b), where the arbitrary function $f(y)$ is in general complex. The amplitude of
the localized solution with arbitrary function of $y$ is now defined by the
equation
\begin{equation}
4\kappa |\rho|^2=2(p+p^*)(1-K)f'_{R}(y). 
\end{equation}
where $f'_{R}(y)$ is the derivative of the real part of $f$ with respect to $y$.  
Thus, the amplitude of the above localized solution varies with the spatial 
coordinate $y$ by virtue of Eq. (26).  This situation is reminiscent 
of the explode-decay dromion of the variable coefficient DSI equation [7] 
where the amplitude varies with time.  It should be mentioned that, to our
knowledge this is the first time the amplitude of a localized solution of a 
(2+1) dimensional nonlinear partial differential equation has been found to 
vary as a function of the spatial coordinate $y$. This can be easily 
generalized to multi-dromions with spatially varying amplitude.

\subsubsection{Induced dromions}
We also wish to point out that the existence of an arbitrary function in the solution 
of $g^{(1)}$ in Eq. (14a) can be further utilized to obtain new induced
dromion solutions [6] for $\chi$.
For example, Eqs. (14a) and (14b) can also be solved in terms of 
arbitrary functions as
\begin{equation}
g^{(1)}=a(y)e^{\hat{\psi_{1}}},\hat{\psi_{1}}=l_{1}x+\omega_{1}t, 
\end{equation}
where $a(y)$ is an arbitrary complex function of $y$ and $l_{1}$ and
$\omega_{1}$ are complex constants constrained by the condition
$i\omega_{1}=l_{1}^2$. Substituting the form (27) into 
(14b) and solving the  resultant equation, we obtain
\begin{equation}
f^{(2)}=b(y)e^{\hat{\psi_{1}}+\hat{\psi_{1}}^*},  
\end{equation}
where $b(y)$ is a real function of $y$ and is given by the condition
\begin{equation}
6\omega_{1R}^2b(y)-l_{1R}b_{y}-8l_{1R}^4b(y) = 2\kappa|a(y)|^2. 
\end{equation}
The above solutions can then be used to generate curved line soliton for the 
field variable and the potential as
\begin{subequations}
\begin{eqnarray}
\chi &=& {a(y) \over {2\sqrt{b(y)}}}e^{i\hat{\psi_{1I}}}\mbox{sech}[\hat{\psi_{1R}}+
{1\over{2}}\mbox{log}b(y)],\\
u &=& 2l_{1R}^2\mbox{sech}^2[\hat{\psi_{1R}}+{1\over{2}}\mbox{log}b(y)]. 
\end{eqnarray}
\end{subequations}
where  $\hat{\psi_{1R}}$ and $\hat{\psi_{1I}}$ are the real and imaginary parts,
repectively of $\hat{\psi_{1}}$.  Thus, by choosing the arbitrary 
functions $a(y)$ and $b(y)$ suitably which are constrained by the Eq. (29), 
one can induce localized solutions  for the field variable $\chi$ as in the case of
Zakharov-Strachan equation [6].  Eventhough there exists two functions $a(y)$ 
and $b(y)$, only one of them is found to be arbitrary, which is evident from the
Eq. (29) above. For example, by choosing
\begin{equation}
{a(y)\over \sqrt{b(y)}} = 2\mbox{sech}(m_{1}y),
\end{equation}
where $m_{1}$ is a real constant, we can find from Eq. (29) that
\begin{equation}
b(y)=\mbox{exp}\left(-\frac{1}{l_{1R}}(8\kappa\mbox{tanh}m_{1}y+(8l_{1R}^4
-6\omega_{1R}^2)y)\right).
\end{equation}
Then we obtain the induced localized solution
\begin{equation}
\chi=e^{i\hat{\psi_{1I}}}\mbox{sech}(m_{1}y)\mbox{sech}\bigg[l_{1R}(x
+2l_{1I}t)-\frac{1}{2l_{1R}}\bigg(8\kappa\mbox{tanh}m_{1}y+(8l_{1R}^4-6\omega_{1R}^2)y
\bigg)\bigg].
\end{equation}

Thus,  by choosing $a(y)$ and $b(y)$ suitably, one can induce a wide class of
localized solutions for the Mel'nikov Eq. (1).
For example, choosing an algebraic form
\begin{equation}
{a(y)\over \sqrt{b(y)}}={2\over{(y+y_{0})^2+1}},
\end{equation}
we obtain an algebraically decaying localized solution
\begin{eqnarray}
\chi &=&{1\over{(y+y_{0})^2+1}}e^{i\hat{\psi_{1I}}}\mbox{sech}\bigg[l_{1R}(x
+2l_{1I}t)-\frac{1}{2l_{1R}}\bigg((8l_{1R}^4-6\omega_{1R}^2)y \nonumber\\
& &
+8\kappa\int
\frac{dy}{((y+y_{0})^2+1)^2}\bigg)\bigg].
\end{eqnarray}
One can as well generalize this procedure to construct even wider class of 
localized solutions.  In fact, multi-induced dromions take the simple form
\begin{equation}
\chi_{N}=\left[{\sum_{j=1}^{N}a_{j}
(y)\over 2\sqrt{b(y)}}\right]
e^{i\hat{\psi_{1I}}}\mbox{sech}[\hat{\psi_{1R}}+{1\over{2}}\mbox{log}b(y)],
\end{equation}
where $a_{j}$, $j=1,2...N$ are arbitrary functions of $y$ and they are related
to $b(y)$ by the relation
\begin{equation}
6\omega_{1R}^2b(y)-l_{1R}b_{y}-8l_{1R}^4b(y) = 2\kappa\sum_{p=1}^{N}
\sum_{q=1}^{N}a_{p}(y)a^*_{q}(y). 
\end{equation}
Note that the sum of the arbitrary functions of $y$ on the right hand side of 
eq.(36) becomes possible due to the structure of eqs. (1). In practice one 
can choose ${\sum_{j=1}^{N}a_{j}(y)\over 2\sqrt{b(y)}}$  conveniently, for 
example, as a combination of algebraic and hyperbolic functions. With N=2, 
choosing the functions as 
\begin{equation}
{a_{1}(y)\over \sqrt{b(y)}} = 2\mbox{sech}(m_{1}y+\delta_{1}),
{a_{2}(y)\over \sqrt{b(y)}} = 2\mbox{sech}(m_{2}y+\delta_{2}),
\end{equation}
where $m_{1}$, $m_{2}$, $\delta_{1}$ and $\delta_{2}$ are parameters, solving eq.(29),
one can obtain
\begin{eqnarray}
b(y)&=&\mbox{exp}\bigg(-\frac{1}{l_{1R}}((8l_{1R}^4-6\omega_{1R}^2)y
+8\kappa(\mbox{tanh}(m_{1}y+\delta_{1})\nonumber \\
&&+\mbox{tanh}(m_{2}y+\delta_{2})+2\int\mbox{sech}(m_{1}y
+\delta_{1})\mbox{sech}(m_{2}y+\delta_{2})dy))\bigg).
\end{eqnarray}	
Then, the induced two dromion solution is given by
\begin{eqnarray}
\chi_{2}&=&e^{i\hat{\psi_{1I}}}(\mbox{sech}(m_{1}y+\delta_{1})+\mbox{sech}(m_{2}y+\delta_{2})) 
\mbox{sech}\bigg[l_{1R}(x+2l_{1I}t) \nonumber \\ 
&&-\frac{1}{2l_{1R}}\bigg((8l_{1R}^4-6\omega_{1R}^2)y 
+8\kappa(\mbox{tanh}(m_{1}y+\delta_{1})\nonumber \\
&&+\mbox{tanh}(m_{2}y+\delta_{2})+2\int\mbox{sech}(m_{1}y
+\delta_{1})\mbox{sech}(m_{2}y+\delta_{2})dy)\bigg)\bigg]
\end{eqnarray}
and is shown in Fig. 2. One can identify the mutual influence of
one dromion over the other from the additional terms occurring in the square
bracket of eq.~(40).
\begin{figure}[!ht]
\begin{center}
\includegraphics[width=0.75\linewidth]{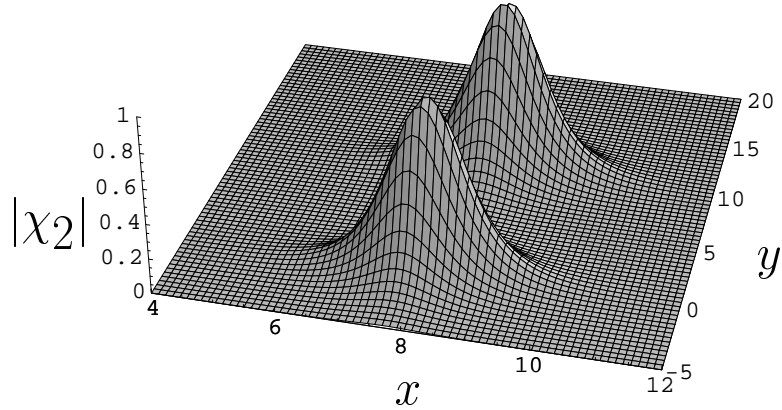}
\end{center}
\caption{Snapshot of the induced two dromion solution of the Mel'nikov 
equation (see Eq. (40))}
\end{figure}

If one chooses both the functions $a_{j}$, $j=1,2$, algebraically, 
\begin{equation}
{a_{1}(y)\over \sqrt{b(y)}} = {2\over{(m_{1}y+y_{10})^2+1}},
{a_{2}(y)\over \sqrt{b(y)}} = {2\over{(m_{2}y+y_{20})^2+1}}
\end{equation}
or one of the functions to be algebraic and the other one to be hypebolic 
\begin{equation}
{a_{1}(y)\over \sqrt{b(y)}} = 2\mbox{sech}(m_{1}y+\delta_{1}),
{a_{2}(y)\over \sqrt{b(y)}} = {2\over{(m_{2}y+y_{20})^2+1}}
\end{equation} 
one can generate different kinds of induced lump-lump or dromion-lump solutions,
respectively.

  We have also tried to generate more general two soliton solutions by choosing
\begin{equation}
g^{(1)}=a_{1}(y)e^{\hat{\psi_{1}}}+a_{2}(y)e^{\hat{\psi_{2}}}
,\hat{\psi_{1}}=l_{1}x+\omega_{1}t, \hat{\psi_{2}}=l_{2}x+\omega_{2}t,
\end{equation}
which lead to a condition $l_{1}=l_{2}$ thereby reducing to our original 
form (36) for $N=2$. This is also true for $N>2$. Thus we believe that the 
solution (36) constitutes the most general localized solution we could 
construct for the Mel'nikov equation through our procedure.

\section{Conclusion}
In this paper, we have pointed out the interesting fact that the (2+1) 
dimensional Mel'nikov equation admits exponentially localized solutions
of different classes. We have also checked its integrability through Painlev\'e analysis.
We have in particular constructed localized dromion solutions and obtained
new classes of localized solutions such as dromions with spatially varying
amplitude and induced dromions.

\section*{Acknowledgement} 
The work of C.S. and M.L. form part of a Department of Science and Technology, 
Govt. of India sponsored research project.


\end{document}